# Preprint Déjà Vu: an FAQ

Paul Ginsparg
Physics and Information Science, Cornell University

This commentary originally appeared in the EMBO Journal.[1] It benefited significantly from editorial supervision, and is reposted here with permission (and minor updates). The introduction provides some overall context.

[1] DOI:10.15252/embj.201695531, 19 Oct 2016. The original title "Déjà Vu all over again" was changed to accommodate European readers unfamiliar with the Yogi Berra allusion.

**Introduction:** Twenty-six years ago, in August 1991, I spent a couple of afternoons at Los Alamos National Laboratory writing some simple software that enabled a small group of physicists to share drafts of their articles via automated email transactions with a central repository. Within a few years, the site migrated to the nascent WorldWideWeb as xxx.lanl.gov (renamed to arXiv.org in 1999) and experienced both expansion in coverage and heavy growth in usage that continues to this day. In 1998, I gave a talk to a group of biologists — including David Lipman, Pat Brown, and Michael Eisen — at a meeting at Cold Spring Harbor Laboratory (CSHL) to describe the sharing of "pre-publication" articles by physicists. The talk was met with some enthusiasm and prompted the "e-biomed" proposal in the following spring (1999) by then NIH director Harold Varmus. He encouraged the creation of an NIH-run electronic archive for all biomedical research articles, including both a preprint server and an archive of published peer-reviewed articles, which generated significant discussion.

I agreed to write a commentary [1] on Varmus' proposal that summer (1999), in part to "comment on some of the attempts in the past half year to isolate physicists, or rather to distinguish their research practices from the rest of the scientific community, in an attempt to assert that what has been so successful and continues to grow 'couldn't possibly' work in say the biological or life sciences."

As I had done in my talk at CSHL a year earlier, I described how arXiv.org had "become a major forum for dissemination of results in physics and mathematics, and suggested some of what we foresee as the advantages of a unified global archive for research in these fields". I also pointed out how it was "entirely scientist driven, and flexible enough either to co-exist with the pre-existing publication system, or help it evolve to something better optimized for researcher needs. In particular, the rapid dissemination ... is not in the least inconsistent with concurrent or post facto peer review, and in the long run provides a possible framework for a far more functional archival structuring of the literature than is provided by current peer review methodology".

I tried further in that commentary to counter the "oft-repeated claim that physics is invariably done in big labs, with large teams, and the papers are all written by hundreds of authors, or that they publish much less", and to counter equally mistaken assertions that physicists are somehow "less competitive" than biologists, or "more interested in hypotheses and less in confirmed results". I mentioned that the way biologists stake intellectual property claims seemed irrational to physicists, leading to the "non sequitur outcome that results could be discussed at a meeting, some other lab could rush to reproduce and rush to publish, and the latter could get full 'credit"'.

I expressed concern that, from the physicists' point of view, "the biologists frequently seem an exceedingly timid group, having ceded direct control over their research results to parties not always acting in their interests".

The "e-biomed" proposal soon morphed into what we now know as PubMedCentral (PMC). Participants M. Eisen and P. Brown from the CSHL meeting together with H. Varmus went on to create the Public Library of Science (PLoS). While neither ultimately had a preprint component, both have played leading roles in the open access movement.

In the summer of 2015, biologist Ron Vale cold emailed me for comments on a draft of an article that described various perceived publication deficiencies in the biological sciences, and intimated that emulating the use of preprints by other fields might help ameliorate some of these [2]. Vale and others soon initiated a movement (ASAPbio) to foster more widespread adoption of preprint usage in biology[2], and I addressed its inaugural meeting in Feb 2016.

Oddly enough, we are still trying to dispel the same misconceptions addressed in my commentary in 1999. I discovered that what I'd written could be used verbatim in 2016, even including the throwaway line, "It's thrilling if the biomedical people are ready to join the 1990s, better late than never ...". The FAQs and answers below are based on questions posed at ASAPbio, and afterwards by Ron Vale, as well as on updates to my late 1990s comments informed by an additional eighteen years of data and experience.

My 1999 commentary closed with "I strongly suspect that, on the one- to two-decade timescale, serious research biologists will also have moved to some form of global unified archive system, without the current partitioning and access restrictions familiar from the paper medium, for the simple reason that it is the best way to communicate knowledge and hence to create new knowledge". My estimate of the timeframe was evidently off, but I am happy to stand by the rest.

**FAQ 1:** Why did you create arXiv if journals already existed? Has it developed as you had expected?

**Answer:** Conventional journals did not start coming online until the mid to late 1990s. I originally envisioned it as an expedient hack, a quick-and-dirty email-transponder written in csh to provide short-term access to electronic versions of preprints until the existing paper distribution system could catch up, within about three months. And it was only for a specific subcommunity, with an estimated hundred articles per year back in 1991. The original intent was to eliminate some of the inadvertent unfairness of the paper preprint distribution, where advance access to information was not uniformly available due to geography or institutional hierarchy.

But the timing was right and it grew rapidly, foreshadowing the emergence of the WorldWideWeb within a few years, followed by the massive societal move to the new electronic communications infrastructure within a decade. In the early 2000s, we did an assay of arXiv usage to see whether the resource had become any less necessary, but the usage was increasing faster than ever. It's now more than a factor of 1,000 larger than the original design capacity, and the submission rate is still growing by about 10% per year.

[2] http://asapbio.org/about-2

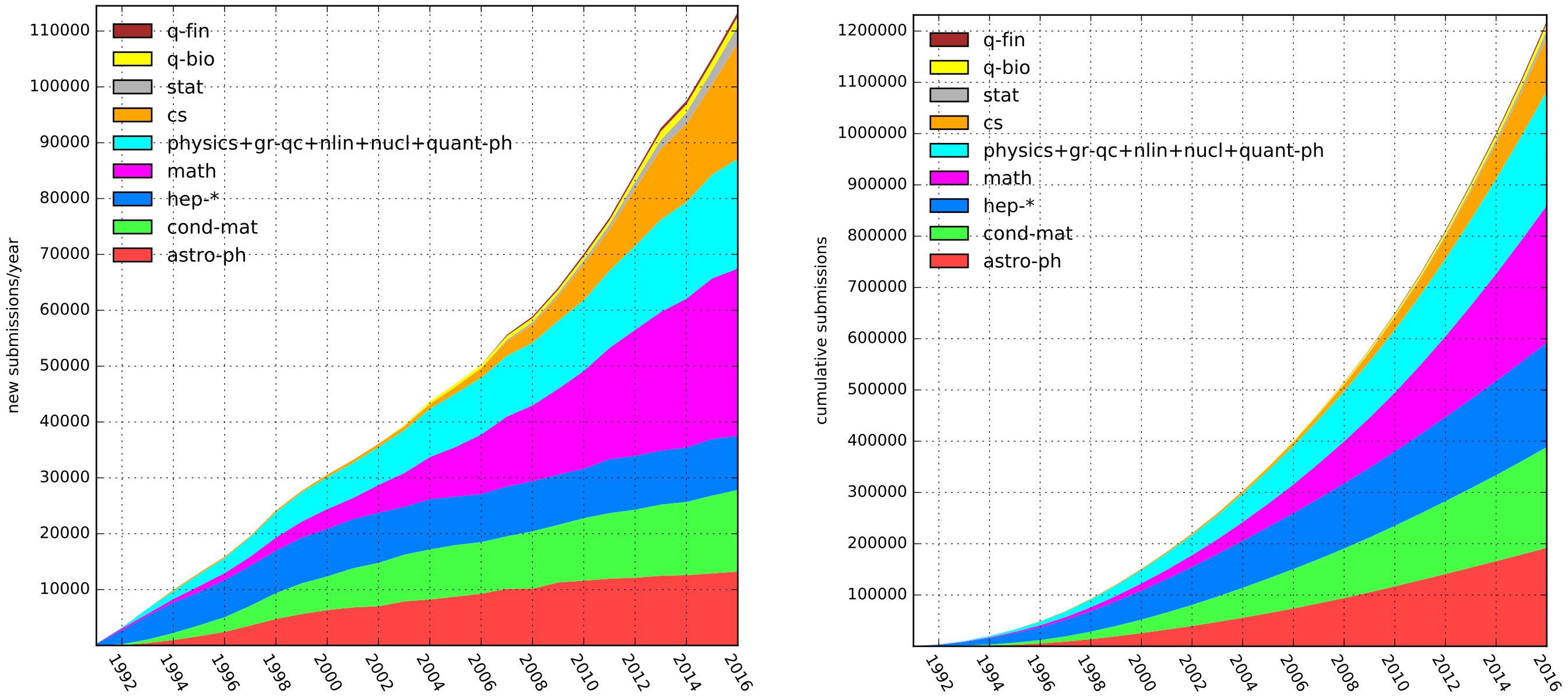

Fig. 1. Yearly and cumulative submissions to arXiv.org, broken out by major subject areas.

**FAQ 2:** How many papers are posted on arXiv per year? How many times are arXiv papers viewed?

**Answer:** arXiv currently holds more than 1.27M articles (as of Jun 2017), with over 120,000 new submissions expected in 2017 (Fig 1). There are over 250,000 active submitters to the system, and that number is currently growing by about 10% per year. The system handles more than 2.5M (non-robotic) accesses per day, including more than 600,000 full-text retrievals, more than 500,000 abstract views, and over 300,000 navigational requests including searches per day. These currently come from more than 300,000 unique visitors per weekday, and over 4M unique visitors per month.

In addition to generic full-text search, we use more sophisticated custom datamining software. The text overlap check processes the 500–1,000 new submissions per day to find duplicates, or any overlap of at least seven words whether by same authors, cited or uncited. We also have a semantic search that goes far beyond what's possible with generic full-text search or standard Google search. The size of the database is a benefit: machine learning tools work much better when trained on millions of articles than thousands, and are still computationally efficient for a linear pass. We also perform a variety of other checks for missing text, missing references, mismatched author names, and of course detecting non-scientific content and other outliers [3]. For many fields of physics, mathematics, and computer science, it is the primary mode of research communication, used for early dissemination, improvement of articles, medium-term visibility, and archival findability.

**FAQ 3:** What are the benefits for scientists to post their work on arXiv?

**Answer:** Posting work on arXiv gives authors a datestamped priority claim, which is accepted by the community, and gives immediate visibility to authors' work. The result has been to speed up the research enterprise, but also to make it fairer, by giving global research communities equal access to the latest results.

**FAQ 4:** Many biologists worry that they will get "scooped" if they place their work on a preprint server. How common is it for someone to see a study posted on arXiv and then try to rush their own paper to a journal to claim credit?

**Answer:** It can't happen, since arXiv postings are accepted as date-stamped and citable priority claims.

Eventually I came to understand that biologists do not use "scoop" in the standard journalistic sense, where it means an exclusive news item of exceptional importance or surprise, with no unethical connotation. Instead, "scooping" in the context of biology research appears to mean the race between laboratories working on overlapping research to get the article published first or, in extreme cases, using information or ideas without proper attribution. The latter is dishonest, of course, regardless of the source. I had long responded that physicists are as or more competitive, in the sense of being eager to be first to discover some new phenomenon and get credit for it.

On the other hand, while fear of unethical behavior may seem widespread in biological circles, it's not clear how prevalent the behavior is in reality, or for that matter *would be* if preprints were widely available. My perception is not that biologists are intrinsically less honest — and I have no reason to suspect it — but rather that they harbor a prevailing fear that *others* will be dishonest. Whether or not founded or exaggerated, the concerns could be addressed by disentangling disclosure of results from their validation. (Vale and Hyman [4] have recently discussed the principles for establishing "priority of discovery", and examine how journal publication dates can obscure priority.)

Systematic posting of preprints, and consensus of the community that they count for staking intellectual precedence, could then provide a long-term solution to the problem. There might be some intermediate pilfering phase, but a few high-profile cases of admonishment and censure would quickly establish a proper ethos. Once preprints achieve higher number, visibility, and easier searchability within a subcommunity, no one can plausibly claim they "did not see it". Biology partitions into subcommunities with sizes ranging from many hundreds into the thousands of researchers, just as in physics and other research areas, so the self-policing mechanisms can be just as effective.

As for concerns that research in biology is fundamentally different from other fields, there are many ideas or calculations in theoretical physics that are much easier to reproduce and claim than would be an experiment in biology. And various tabletop experiments in condensed matter physics might be roughly comparable to those in biology in that regard.

But the experience has been that unexpected or rapid progress leads to increased preprint usage within communities, precisely to stake priority claims, and that increased usage remains the norm afterward.

**FAQ 5:** Large-group effort experimental physics (e.g., in particle physics) and theoretical physics are very different from biological research. Do small experimental physics groups also use arXiv?

**Answer:** First, and perhaps contrary to popular perception, it's worth recalling that only a very tiny percentage of the world's physicists work in large experimental groups. Rather than a monolithic community, "physics" is a mixture of very different cultures, from low-temperature experiment, to astrophysics, to biophysics, to quantum information, and a wide variety of others.

And yes, small experimental physics groups also use arXiv. Condensed matter experimentalists, for example — researchers working on superconductors, superfluids, liquid crystals, polymers, quantum information, graphene to semiconductors — regularly upload their work. They began posting more slowly than theorists, both to benefit from the additional time to publication to do follow-up work and to receive referee feedback, without reputational risk. Perhaps they are more like biologists in that sense.

But, as mentioned above, every once in a while an area goes hot. In 1987, four years prior to the start of arXiv, a "high Tc superconductor" craze began when materials made from ceramics were discovered to remain superconducting at temperatures higher than previously seen. Experimental results were shared via nth-order photocopies of faxed preprints, high technology at the time. After arXiv started, I remarked how it could have grown that much more quickly in those areas had those events occurred only a few years later. Happily, there were a few later such events, including interest in the magnesium diboride superconductors in 2001 and the iron pnictide superconductors starting in 2008. Each time the associated experimental communities used arXiv to report breaking results and stake precedence claims, and then remained as dedicated users. So far, no community that has adopted arXiv for rapid dissemination has since abandoned it.

**FAQ 6:** In competitive areas, is there a race to post preprints, resulting in a decrease in the quality of communicating scientific work?

**Answer:** Serious researchers typically take the utmost care before submitting to arXiv, precisely because the work will be exposed to the entire world, and naive errors would be both highly embarrassing and by design not removable. For this reason, some claim to "sweat bullets" before hitting the submit button. There is presumably a background of careless authors, but they're as ignorable in competitive areas as elsewhere.

There is also the question of staking claims prematurely. Fortunately, there is less ambiguity in assessing these on arXiv than in other fora: Either all the details are in place and date-stamped, or they're not. If the details are added later, or if the claim subtly shifts, that too can be retroactively adjudicated at ease, without having to rely on faulty memories. Authors who repeatedly make bold claims only to withdraw them quickly lose credibility. Historic data on accesses to submissions in their first few days can be correlated with the number of citations accumulated after many years and is very predictive [5]. There is also a retrieval signature for articles that promise more in the abstract than delivered in full text, and it anti-correlates with activity levels of subsequent articles by the same submitter.

**FAQ 7:** Do grant committees, prize committees, and university promotion committees consider arXiv preprints in their decision-making processes?

**Answer:** There are no fixed policies, and I do not have extensive data, but I can speak from my own experience. Preprints are certainly used as evidence of recent productivity in applicant CVs for jobs and grants. Searching successful grant applications online at the NSF site, for example, provides many examples of citations to preprints on arXiv. A reviewer for a grant application in a field that heavily uses arXiv would likely be surprised to see no recent relevant work posted there, and that could enter explicitly or implicitly into considerations.

Similarly, in initial hiring, we try to assess the trend of a candidate's career trajectory, and, all else being equal, their preprint record as a measure of productivity over the past couple of years works much better as a predictor of productivity over the next few years than older journal publications. Of course, this requires some expert assessment, but most departments have people in house who can assess a candidate's preprints as well as anonymous referees or journal editors. In the later stages, say for tenure, candidates would be expected to have journal publications as well, since work done with students and postdocs is ordinarily submitted for peer review.

References to preprints are regularly used by institutions in their press releases of new prominent work, and they are then carried as well by the conventional news media, blogs, and social media.

Regarding prize committees, there is at least one prominent example of having considered arXiv preprints: Grigori Perelman was awarded both the Fields Medal in 2006 and the $1M Millennium prize in 2010 for his proof of the Poincare conjecture in three dimensions, which appeared in 2003 only on arXiv. It's difficult to imagine why results accepted and used by an entire expert community would be disqualified for consideration because they didn't pass through a less stringent journal filter.

**FAQ 8:** Without a formal peer-review process, has "pseudo-scientific" work (for example by politically motivated individual opposed to climate change) slipped into arXiv and has this proven to be a problem?

**Answer:** Some has probably slipped in, on the basis of coming from institutional authors with an otherwise conventional publication record; others might have gone unnoticed. But it has not proven to be problematic. The site is known to have a loose form of moderation and a very occasional problematic item slipping through is the price for rapid dissemination of the rest. In addition, back-and-forth over a contentious submission from an otherwise legitimate submitter can force corrective revisions. In any event, many attempts per week are blocked by moderators, so that the site does not become a platform for non-scientists to employ as a megaphone. If an article that is later found to be fraudulent is posted, the current version can be replaced by site administrators with a withdrawal notice, explaining the reason — as has happened a handful of times.

The quality control employed by arXiv is unique: not uniquely creative by any means, but unique in its implementation of employing a large group of human moderators (active scientists) to glance at incoming submissions and judge the appropriateness for the subject area — usually based just on title/abstract — and for being above some minimal bar of plausible interest to the research community[3]. Sometimes the process works better than journal review, for instance when moderators work above and beyond the call of duty to spare ill-advised graduate students unnecessary embarrassment (not that it results in much gratitude [6]).

As arXiv continues to grow in prominence, the stakes become higher as it plays an increasing role interfacing with journalists, and with the general public. It also operates on an unforgiving daily turnaround, so in recent years the human moderation has been supplemented by an automated machine learning framework which can flag and hold potentially problematic submissions for additional human scrutiny.

Some of the decisions moderators make are subjective, for instance whether non-scientific manuscripts such as literature reviews, commentaries, articles about science policy, historical accounts, or short conference submissions will be of interest to the research community. There are some general guidelines given to promote consistency across subject areas, but there are inevitably some minor policy differences between moderators of different subject areas. There is also a formal appeal procedure, adjudicated by designated appellate moderators, and as well overseen by the subject area chairs who ultimately report to the scientific advisory board.

Moderators are aware that the media might jump on questionable claims, though the experience has been that journalists are well aware of the pitfalls of a preprint site, and do a good job of both getting expert feedback before proceeding with coverage and qualifying the nature of the source to readers. We know from the usage statistics which submissions get large amounts of public attention, amplified by a fascinating interplay between newspapers, blogs, and social media.

[3] http://arxiv.org/help/moderation

This is not to claim, however, that this minimal form of quality control is remotely sufficient. In an imaginary world, in which electronic preprints were invented first, we'd still have had to create something like journals, though perhaps with a different implementation of the peer-review methodology.

**FAQ 9:** What happens when incorrect work gets posted on preprints?

**Answer:** Authors always have the option of submitting revised versions, with corrections, or they can post a withdrawal notice with explanation for the action taken. In either case, all previous versions remain archived and accessible for comparison, with their original date-stamps. (See also FAQ 6 and FAQ 16.)

Moderators could certainly force retraction or correction, though in practice it is usually readers who notice that something is amiss. Sometimes co-authors complain that an article wasn't cleared for submission, sometimes people insist they should have been co-authored, sometimes that they shouldn't have been co-authored, or someone objects to fraudulent use of affiliation. Moderators are sometimes enlisted to check past submissions to assess whether there's a pattern. Overall, moderator-induced retractions and corrections are very rare — less than 0.1% of submissions — and they've been rarer still in recent years with the use of the above-mentioned automated checks (FAQ 2).

**FAQ 10:** Do all physicists and mathematicians use arXiv? If not, why not?

**Answer:** Any physicist or mathematician who uses an Internet search engine will likely end up reading articles on arXiv, since so many searches lead to it. arXiv receives a continuously increasing fraction of the articles ultimately published in these fields, but it is not 100%, so there remain researchers who don't systematically upload articles. This could be a function of age (younger researchers typically do submit everything they write), or more likely a question of a research subcommunity that has not yet fully accommodated to arXiv usage. Once some critical mass of prominent researchers in a community adopts it for research communication, then the vast majority of the community quickly follows.

**FAQ 11:** Are there physics or mathematics journals that will not accept a manuscript for review if it has been previously posted as a preprint?

**Answer:** I can't think of any offhand. Recall that arXiv was a fait accompli before any journals were online. Authors had established their clear preference to continue using it, and journals cannot risk alienating their authors.

**FAQ 12:** What are the biggest issues or tensions between arXiv and the journals?

**Answer:** From arXiv's standpoint, there are few if any. The American Physical Society, a major physics publisher but also a professional society, set the tone in the late 1990s by accepting practices already adopted by its membership and endorsing arXiv usage. arXiv screens for copyright violations, and during the submission process submitters confirm that they have the right to make the deposition, and assign to arXiv a non-exclusive license to distribute. Under copyright law, such a license takes precedence over any subsequent copyright reassignment, so unless submitters are somehow confused there aren't any legal issues.

Many journals go out of their way to make it simpler for authors to submit, just by specifying the arXiv identifier, so that the journal can review the arXiv version. This has been simple for journals to automate, since the arXiv URLs for abstracts and full texts are all simply specified by the arXiv id. Some journals even permit depositing the journal-created pdf, and all permit authors to continue updating the author-created version on arXiv. (arXiv ids long predate DOIs, so have proven at least as archivally stable.)

**FAQ 13:** When do scientists post on arXiv? Prior to, at the same time or after journal peer review?

**Answer:** All of the above can happen. Some authors post somewhat before submitting to a journal, some post simultaneously, some post sometime in the middle of the review process, some wait until it has been accepted by a journal, some wait until it has officially appeared in the journal, and some wait until long after that (journal policies permitting).

**FAQ 14:** How common is it for physicists and mathematicians to submit work to a traditional journal after posting on arXiv?

**Answer:** I recently looked at submissions from 2007 to 2014 in high-energy physics, where we have bibliographic services that systematically match arXiv identifiers to journal references. (In all fields, authors are encouraged to add a journal reference and DOI when available, but don't always remember to do so.) More than 80% of those articles were published in traditional journals, and it is important to note that the vast majority of the rest were items such as conference proceedings, theses, and lecture notes, not intended for journal publication but nonetheless also subject to some form of review.

**FAQ 15:** Does arXiv stimulate dialogue that helps to correct or improve work before journal publication?

**Answer:** Yes, authors benefit greatly from feedback from interested readers, contributing to improved versions of articles, which are then uploaded to arXiv as new versions. This is important, since later versions of articles that are simultaneously submitted to journals can benefit from both the journal-mediated peer review and the "crowdsourced" review. In other words, both the published version and the final arXiv version can benefit from suggestions via channels other than, or in addition to, journal-mediated peer review. Sometimes additional suggestions come after the "definitive" journal version is published, in which case the final updated arXiv version can be even more useful to readers.

While the original intent of the preprint server was rapid dissemination, it very quickly became the go-to place for archival access as well, and this an important component of its utility and popularity. Authors are understandably determined to propagate correct information whenever possible, so rather than let readers be misinformed or confused, they typically make immediate corrections to a latest arXiv version, since that's what many readers access, either before or after publication elsewhere. This is the inevitable consequence if preprint servers come to be regularly used for archival access.

**FAQ 16:** Can you explain the "version" system on arXiv and why you use it?

**Answer:** From the outset, submissions to arXiv were assigned a persistent identifier, and it was possible to update the submission while retaining the identifier. The problem was that added content could then be inadvertently backdated, or content later corrected or removed could leave dangling citations. To eliminate all such issues, arXiv retains and makes available all successive date-stamped versions — just as Wikipedia later implemented — where the first version has v1 appended to the identifier, second has v2, and so on. It is possible to link directly to any given version, and in the absence of a specific version number, the default is to serve up the latest version.

arXiv identifiers are currently of the form arXiv:1607.12345, where the first four digits give the last two digits of the year and the month, and the digits after the period give the accession number within the month. When cited in that form, the arXiv provenance is clear, and it is clear that it refers to material on a preprint server, not necessarily peer-reviewed. This is how arXiv content has been cited by other articles, blogs, social media, and journalists for decades.

**FAQ 17:** Why does arXiv not have a commentary section on scientific work? How do scientists exchange ideas in response to a preprint?

**Answer:** Well over 20 years ago, we had considered comments and numerical rating systems, but received unambiguous feedback to remain focused on the basic task of dissemination. Part of the issue was that contentious comments would have to be moderated, since the system already played such a prominent role for the research community. This in turn would have required human labor, but the scalability of the system depended on automating as much as possible, so ultimately could have both distracted and detracted from the primary mission. The decision then was that such facilities could and should operate at a logical (and physical) remove from the main site archiving functions, and it remains a good one.

It's worse than just ambivalence about comment threads, however. Unlike fields with no universal dissemination system, where users might all claim to be in favor in principle (but not participate in practice), arXiv has very vocal users who are not just mildly negative about comment threads, but *adamantly opposed* to having them mediated via the main site. This attitude was recently reinforced by a broad user survey. Authors regard the drama-free minimalist dissemination as a prominent virtue, which contributes to arXiv's success.

About a decade ago, when blogging became popular, we experimented with "track-backs" linked from the abstract pages, and that provided a distributed means of moderating the discussions. Researchers also have the option of posting formal comments in response to submissions and regularly communicate in "non-public channels" via email. For the time being, it makes sense to continue to piggyback on existing external services, rather than adding a possibly faulty new wheel.

**FAQ 18:** What is the relationship of arXiv to blogs and social media?

**Answer:** arXiv has no special relationship to blogs and social media, other than providing a stable and transparent link structure for them to direct readers to abstracts and full texts. Blogs and social media occasionally refer large numbers of readers to arXiv articles and can provide a source of useful commentary. Some of these are linked back from the "trackbacks" link on the arXiv abstract page. It should soon be possible for authors to curate their own set of links directly from the abstract pages to these and other useful external resources.

**FAQ 19:** How is arXiv funded and governed?

**Answer:** In 2012, arXiv adopted a model[4] in which it is collaboratively governed and supported by the research communities and institutions that benefit from it most directly. It is currently supported by funds from a network of member libraries, the Simon's foundation and financial support, labor, and infrastructure provided by the Cornell University Library. The annual membership fees depend on the institutional usage and range from $1,500 to just $3,000 per year, comparable to the author fees for a single open access article.

According to governance principles[5], the "Cornell University Library holds the overall responsibility for arXiv's operation and development, with strategic and operational guidance from its Member Advisory Board (MAB) and its Scientific Advisory Board (SAB)".

[4] https://arxiv.org/help/support

[5] https://confluence.cornell.edu/download/attachments/127116484/arXivPrinciplesMarch12.pdf

The MAB is a small group of representatives from nearly 200 libraries in 24 countries, who have made contributions to support arXiv via the membership program, and the SAB is a small group of researchers representing the interests of the user community. The chairs of the moderation boards for the largest subject areas — physics, mathematics, and computer science — are *ex officio* members of the SAB, as is a representative from the Simon's Foundation, and as am I. The MAB and SAB communicate independent of one another via email, periodic conference calls, and annually in-person meetings. Members of the SAB serve renewable three-year terms and also provide oversight to the moderation boards by setting policy and serving as the final appeal level.

**FAQ 20:** What are the key ingredients that you feel have been important for the success of arXiv?

**Answer:** A key ingredient is "single-stop shopping": Once it achieved critical mass within a research community, readers knew they could use it as their primary, and perhaps even only, information feed, with no arbitrary partitioning of the literature into different databases. It also helped that the bibliographic data were carried by indexing services already used by researchers. It was important that the content was held centrally rather than via links to distributed resources, to ensure uniform formatting, familiar interface, universal accessibility, and especially to ensure archival stability. It is possible that an equally effective system could now be created from distributed articles using a centralized index and aggregator, but decades ago that would not have been sufficiently robust.

The growth into new subject areas was always via a bottom-up "grass roots" approach, mirroring innate human tribalism. Active and prominent members of small subcommunities began using it, building up to critical mass, and then it quickly saturated. Cross-disciplinary researchers became aware of it as a good way to share research information and seeded the next subcommunity, and so on.

**FAQ 21:** What are the key messages from the recent arXiv user survey?

**Answer:** A survey of arXiv users was conducted in spring 2016 by the Cornell University library to help assess current policies and prioritize development work [7]. It was also to determine user reaction to some of the automated tools for checking submissions that were put online within the past few years. There were close to 40,000 respondents, well distributed with respect to career status (faculty, staff, postdoc, graduate students), subject areas of interest, age (70% were under 40), and years of using arXiv (25% for 11 or more years). In total, 99% were overall satisfied, slightly better than I receive in my biannual teaching evaluations.

I've already mentioned the opposition to having comment threads directly on the main site: More generally users weighed in to keep things simple and came out strongly against trying to evolve arXiv into any form of social network. Users would not object to improved search facilities, and also prioritized "better support for submitting and linking research data, code, slides and other materials associated with papers". This will expedite the deployment of the above-mentioned facility for active curation of links by submitters, which will improve interoperability with a variety of external services (including the possibility for new overlay journals or for semantic overlays to gain visibility and traction).

And it was heartening to hear that "users were in agreement about the importance of continuing to implement quality control measures, such as checking for text overlap [8], correct classification of submissions, rejection of papers without much scientific value, and asking authors to fix format-related problems", since these were some of the automated tools that I had put online.